\documentclass[aps,prl,twocolumn,groupedaddress]{revtex4-1}
\usepackage{graphicx}

\begin{document}


\title{Constraining the spin-independent elastic scattering cross section of dark matter using the Moon as a detection target and the background neutrino data}


\author{Man Ho Chan, Chak Man Lee}
\affiliation{The Education University of Hong Kong, Tai Po, Hong Kong, China}


\date{\today}

\begin{abstract}
Our Moon is a natural giant direct-detection target for constraining dark matter. By considering the dark matter capture rate of the Moon, we obtain some constraints of the spin-independent elastic scattering cross section of dark matter particles on nucleons $\sigma_p^{\rm SI}$ using the background neutrino data. The upper limits of $\sigma_p^{\rm SI}$ can be constrained to $\sim 10^{-38}-10^{-36}$ cm$^2$ for certain `resonance dark matter mass' ranges. These stringent astrophysical constraints are complementary to the constraints obtained by the direct-detection experiments.
\end{abstract}

\maketitle


\section{Introduction}
It is commonly believed that some unknown dark matter particles exist in our universe to account for the missing mass. Although we do not have any ideas on what dark matter particles are, many direct- and indirect-detection experiments are going to search for the signal of particle dark matter. Based on these detections, a large parameter space of dark matter properties has been constrained \cite{Ackermann,Chan2,Abecrcrombie,Akerib,Aprile,Chan}.

Beside the man-made detectors, our Moon is a natural giant direct-detection target for constraining astrophysical theories. For example, some recent experiments are searching for the radio signals produced from the interactions between ultrahigh-energy cosmic ray particles and the Moon \cite{Veen,Aminaei}. These results can help us understand more about high-energy astrophysics. Moreover, some studies are focusing on the dark matter capture rate by the Moon \cite{Garani}. By considering the heat production rate inside the Moon due to dark matter annihilation, one can obtain some constraints of the spin-independent elastic scattering cross section of dark matter particles on nucleons $\sigma_p^{\rm SI}$ \cite{Garani}. Similar analyses have been done by using different astrophysical objects, such as white dwarfs \cite{McCullough}, the Sun \cite{Silk} and Mars \cite{Bramante}. These astrophysical constraints are complementary to the constraints obtained by the direct-detection experiments.

In this article, using the Moon as a detection target, we discuss a new analysis with neutrino data to constrain $\sigma_p^{\rm SI}$. We show that this method can give tighter astrophysical constraints than that obtained by previous studies using the argument of energy or heat release for certain ranges of dark matter mass, especially for the neutrino annihilation channels. 

\section{Formalism for the dark matter particle capture model}
The time evolution of the dark matter particles gravitationally captured by an astrophysical object, using the Moon as an example for our current study, is given by
\begin{equation}
\frac{{\rm d}N_{\chi}}{{\rm d}t}=C-AN_{\chi}^2-FN_{\chi},
\label{dNdt}
\end{equation}
where $C$ is the dark matter capture rate due to the dark matter-nucleon interactions, whereas $AN_{\chi}^2$ and  $FN_{\chi}$ govern the number of dark matter particles ($N_{\chi}$) lost due to their annihilation and evaporation, respectively.

The dark matter particles inside our Galaxy could scatter off the Moon via collisions. Eventually, they will lose energy so that their resulting velocities are lower than the Moon's escape velocity at a distance from the center of the Moon, $v_e(r)$, and will finally be trapped in the Moon. The capture rate $C$ depends on the Moon's composition, which consists of several species with number densities $n_i(r)$ in three different layers. Two benchmark models, named MAX and MIN, are considered in this analysis. The MAX and MIN models respectively correspond to maximal and minimal core density models \cite{Garani}. Generally speaking, the MIN model would be a more conservative model for constraining dark matter. The corresponding information for these two models is shown in Table 1 and Table 2. The capture rate of dark matter particles of mass $m_{\chi}$ with the scattering cross section $(\sigma_i)$ of dark matter on nuclei of type $i$ is then given by \cite{Garani,Gould1,Gould2}
\begin{eqnarray}
C&=&\sum_{i}\sqrt{\frac{6}{\pi}}\frac{\rho_{\rm DM}}{m_{\chi}v_d}\int dV v_e^2 \sigma_i n_i\times\nonumber\\
&&\frac{1}{2\eta A_i^2}\left[\left(A_{i,+}A_{i,-}-\frac{1}{2}\right)(\chi(-\eta,\eta)-\chi(A_{i,-},A_{i,+}))\right.\nonumber\\
&&\left.+\frac{1}{2}A_{i,+}\exp(-A_{i,-}^2)-\frac{1}{2}A_{i,-}\exp(-A_{i,+}^2)\right.\nonumber\\
&&\left.-\eta\exp(-\eta^2)
\right]
,\label{capture}
\end{eqnarray}
where $\rho_{DM} = 0.3~{\rm GeV~cm^{-3}}$ is the local dark matter density, $\chi(a,b)=(\pi/2)[{\rm Erf}(b)-{\rm Erf}(a)]$, $\eta=3u^2/2v_d^2$ is the dimensionless velocity and 
\begin{eqnarray}
A_i^2&=&6\frac{v_e^2}{v_d^2}\frac{m_{\chi}m_i}{(m_{\chi}-m_i)^2}, \\
{\rm with }&&\nonumber\\
A_{i,\pm}&=&A_i\pm\eta.\nonumber  
\end{eqnarray}
Here $u=220$ km/s is the velocity of the Moon, and the dark matter distribution is assumed to be Maxwellian with the velocity dispersion $v_d = 270~{\rm kms^{-1}}$. The terms involving $A_{i,\pm}$ are responsible for a composition-dependent resonance-like behavior of the capture rate, which gives sharp peaks at some narrow ranges of dark matter mass $m_{\chi}$. We integrate Eq.~(2) over all three layers containing different elemental compositions to get the total capture rate $C$.

We specify the cross section of dark matter on nucleons to be spin-independent elastic scattering. Assuming isospin equivalence, for a nucleus with mass number $\tilde{A}_i$ and charge $Z_i$, the dark matter-nucleus spin-independent elastic scattering cross section in Eq.~(\ref{capture}) can be expressed in terms of that of proton $(\sigma_p^{SI})$ by \cite{Garani}
\begin{equation}
\sigma_i^{\rm SI}=\left(\frac{\mu_r^{\tilde{A}_i}}{\mu_r^{p}} \right)^2 \tilde{A}_i^2\sigma_p^{\rm SI},
\end{equation}
where $\mu_r^p=m_{\chi}m_p/(m_{\chi}+m_p)$ and $\mu_r^{\tilde{A}_i
}=m_{\chi}m_{\tilde{A}_i}/(m_{\chi}+m_{\tilde{A}_i})$ are the reduced masses of proton-dark matter and nucleus-dark matter, respectively. The capture rate as a function of dark matter mass is shown in Fig.~1. We can see that the capture rate is very large when $m_{\chi}$ is equal to some `resonance dark matter mass' values (e.g. $m_{\chi} \approx 37$ GeV or $m_{\chi} \approx 52$ GeV).

\section{Neutrino flux emission due to dark matter annihilation}
Generally speaking, the captured dark matter inside the Moon would undergo annihilation to give a large amount of photons, electrons, positrons and neutrinos. The emitted photons and electron-positron pairs would interact with the Moon to contribute internal heat while the emitted neutrinos would almost completely escape from the Moon. Some of the neutrino flux would pass through the Earth and contribute to the atmospheric neutrino flux.

The coefficient $A$ in Eq.~(1) can be approximately given by \cite{Garani}
\begin{equation}
A \approx \frac{\sigma v}{V},
\end{equation}
where $V$ is the thermal volume and $\sigma v$ is the annihilation cross section. Ref.~\cite{Garani} shows that dark matter could be thermalized by the lunar matter so that the dark matter temperature is close to the Moon core temperature $T_{\rm core}=1700$ K. The thermal radius and volume are given by $R_{th}=[9T_{\rm core}/(4\pi G\rho_{\rm core}m_{\chi})]^{1/2} \approx (4.0-5.6) \times 10^3({\rm GeV}/m_{\chi})$ km and $V=(4\pi/3)R_{th}^3$ respectively, where $\rho_{\rm core}$ is the core density of the Moon \cite{Garani}. Here, we can see that for $m_{\chi}>5$ GeV, the captured dark matter cloud is completely inside the Moon.

In the followings, we will consider the electron neutrino flux only as it can give the tightest constraints for dark matter. The electron neutrino flux density emitted due to dark matter annihilation, with sufficiently large distance $D_L \approx 3.84\times 10^5$ km between the Moon and the Earth, is expressed by
\begin{equation}
\Phi_{DM}=\frac{1}{4\pi D_L^2}\left(AN_{\chi}^2\right)\sum_{i=1}^3 \left[f_i\frac{{\rm d}N_{\nu_i,{\rm inj}}(E,m_{\chi})}{{\rm d}E}\right],
\label{SDM}
\end{equation}
where ${{\rm d}N_{\nu_i,{\rm inj}}(E,m_{\chi})}/{{\rm d}E}$ is the injected energy spectrum of dark matter annihilation contributed by a certain type of neutrino $\nu_i$ \cite{Cirelli}. Note that neutrinos can oscillate and change to different flavors during transmission. Based on the observed mixing angles $\theta_{13}=8.5^{\circ}$, $\theta_{23}=33^{\circ}$ and $\theta_{12}=42^{\circ}$ \cite{Fantini}, approximately 80\% of the injected electron neutrinos, 10\% of the injected muon neutrinos and 10\% of the injected tau neutrinos would become electron neutrinos after traveling a distance of $D_L$. Therefore, we take $f_1=0.8$ and $f_2=f_3=0.1$. 

After a few billion years of dark matter capture, the number of dark matter particles in the dark matter cloud would approach an equilibrium (i.e. ${\rm d} N_{\chi}/{\rm d}t =0$). Therefore, the capture rate is equal to the sum of the annihilation rate and evaporation rate. Generally speaking, the evaporation rate is important only for certain ranges of $m_{\chi}$ and $\sigma_p^{SI}$ (see Fig.~2). The calculation details of the evaporation rate can be found in \cite{Garani}. For the parameter space in which evaporation is not important, the annihilation rate $AN_{\chi}^2$ in Eq.~(\ref{SDM}) can be simply replaced by the capture rate in Eq.~(\ref{capture}).

The neutrino flux (including electron neutrinos, muon neutrinos and tau neutrinos) emitted from dark matter annihilation would contribute to the background electron neutrino flux reaching the Earth. The amount of GeV background electron neutrino flux can be estimated from the observed atmospheric neutrino flux \cite{Honda,Daum}. Based on the observational data of the atmospheric neutrino flux \cite{Gaisser}, the background electron neutrino flux of $7-100$ GeV can be parameterized as $\Phi_{\rm back}=0.014\times E_{\nu,~{\rm GeV}}^{-3.5}~ {\rm GeV^{-1}cm^{-2}s^{-1}sr^{-1}}$. The calculated flux is based on the calculation model in \cite{Honda} and the neutrino data are obtained from the Super Kamiokande detector in Japan. This detector has made many important measurements including precise detection of the solar neutrino flux and atmospheric neutrino flux \cite{Gaisser}. Since $\Phi_{DM}$ must be less than $\Phi_{\rm back}$, we can find the upper limits of $\sigma_p^{\rm SI}$ for different dark matter mass and annihilation channels by setting $\Phi_{DM}=\Phi_{\rm back}$. 

In Fig.~2, we show the corresponding upper limits of $\sigma_p^{\rm SI}$ as a function of dark matter mass for four popular annihilation channels ($e^+e^-$, $\mu^+\mu^-$, $\tau^+\tau^-$ and $b\bar{b}$ channels). In particular, the upper limits for the $\mu^+\mu^-$, $\tau^+\tau^-$ and $b\bar{b}$ channels are generally tighter than that obtained by considering the internal heat rate of the Moon (the Moon internal heat constraint) \cite{Garani}. For a small range of the `resonance dark matter mass' near 52 GeV ($m_{\chi} \approx 48-56$ GeV), the upper limits can be constrained to $\sigma_p^{\rm SI} \sim 10^{-36}$ cm$^2$, which are tighter than that obtained by considering the internal heat rate of the Earth (the Earth internal heat constraint) \cite{Mack}. Furthermore, in Fig.~3, we show the upper limits of $\sigma_p^{\rm SI}$ for three neutrino annihilation channels. These limits are even constrained to $\sigma_p^{\rm SI} \sim 10^{-38}$ cm$^2$, which are much tighter by at least two orders of magnitude. For $m \ge 35$ GeV, our constraints for neutrino annihilation channels are tighter than the previous constraints. We also include some of the current direct-detection upper limits (XENON1T, LUX and DEAP-3600) \cite{Aprile,Akerib,Yaguna} in Fig.~3 for reference. Although the current direct-detection bounds are stronger than our limits, our study can provide an alternative and complementary analysis for constraining the interaction between dark matter and ordinary matter. 

Note that there is an assumption that dark matter would interact with quarks in order for it to be captured in the Moon. Therefore, it may be impossible for the captured dark matter to self-annihilate to give 100\% primary neutrinos (i.e. 100\% via the neutrino annihilation channels). Nevertheless, the limits for the neutrino channels in Fig.~3 can show an extreme limit to benchmark the experimental reach of the proposed analysis. Moreover, the calculations of the Moon internal heat constraints in \cite{Garani} and the calculations of the Earth internal heat constraint in \cite{Mack} have assumed that all energy released in dark matter annihilation contributes to the internal heat of the Moon or the Earth. However, for $\mu^+\mu^-$, $\tau^+\tau^-$ and $b\bar{b}$ channels, nearly half of the energy released in dark matter annihilation is in the form of neutrino emission \cite{Cirelli}, which would finally escape from the Moon and the Earth. For the neutrino annihilation channels, nearly 85\% of the energy is in the form of neutrino emission \cite{Cirelli}. The Moon and the Earth internal heat constraints shown in Fig.~2 and Fig.~3 have been revised by this consideration. 

\begin{figure}
 \includegraphics[width=80mm]{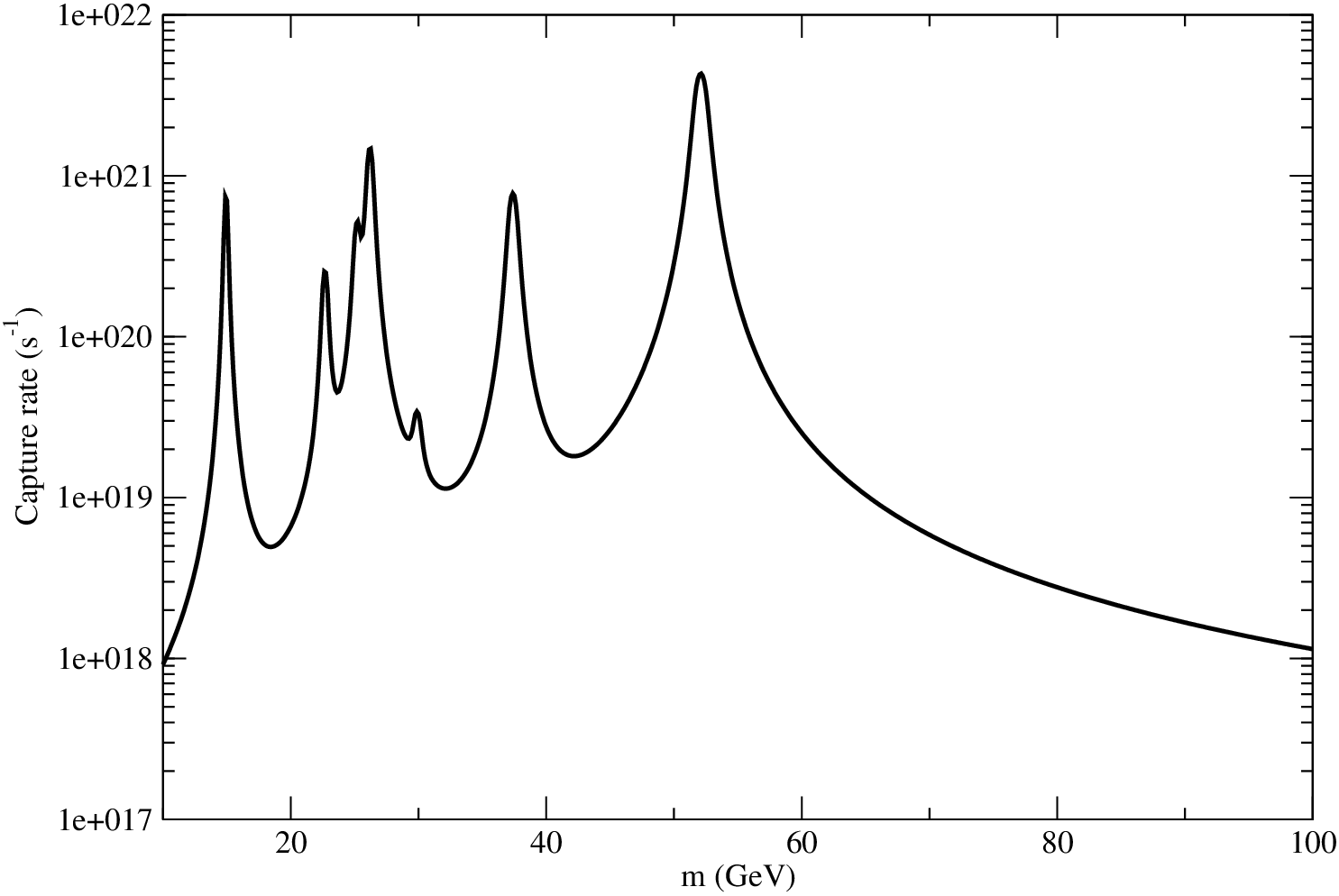} \newline
 \vskip 8mm
 \includegraphics[width=80mm]{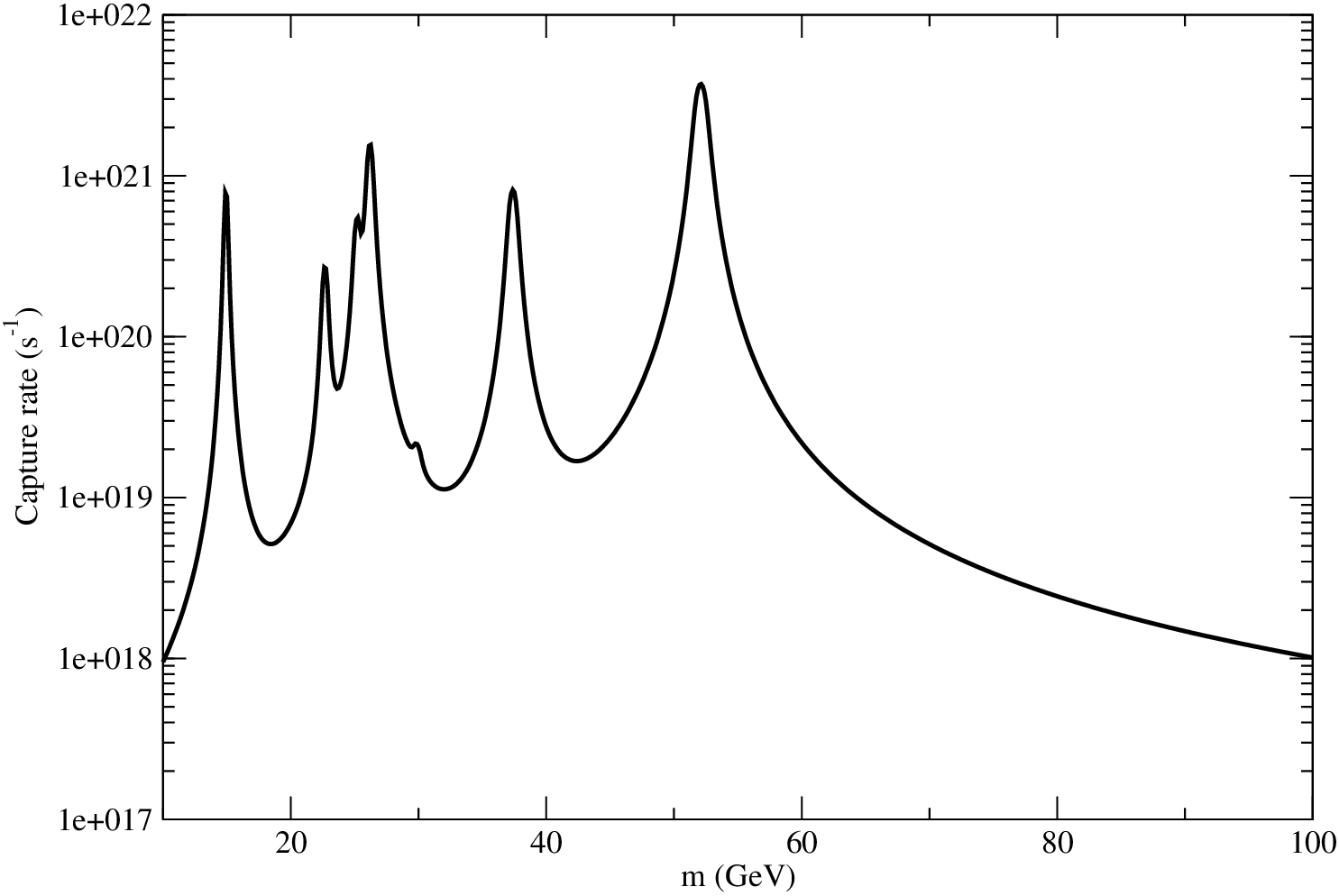} 
\caption{The capture rate of dark matter particles, assumed $\sigma_p^{\rm SI}=10^{-33}$ cm$^2$ for MAX (upper figure) and MIN (lower figure) models.}
\label{Fig1}
\end{figure}

\begin{figure}
 \includegraphics[width=80mm]{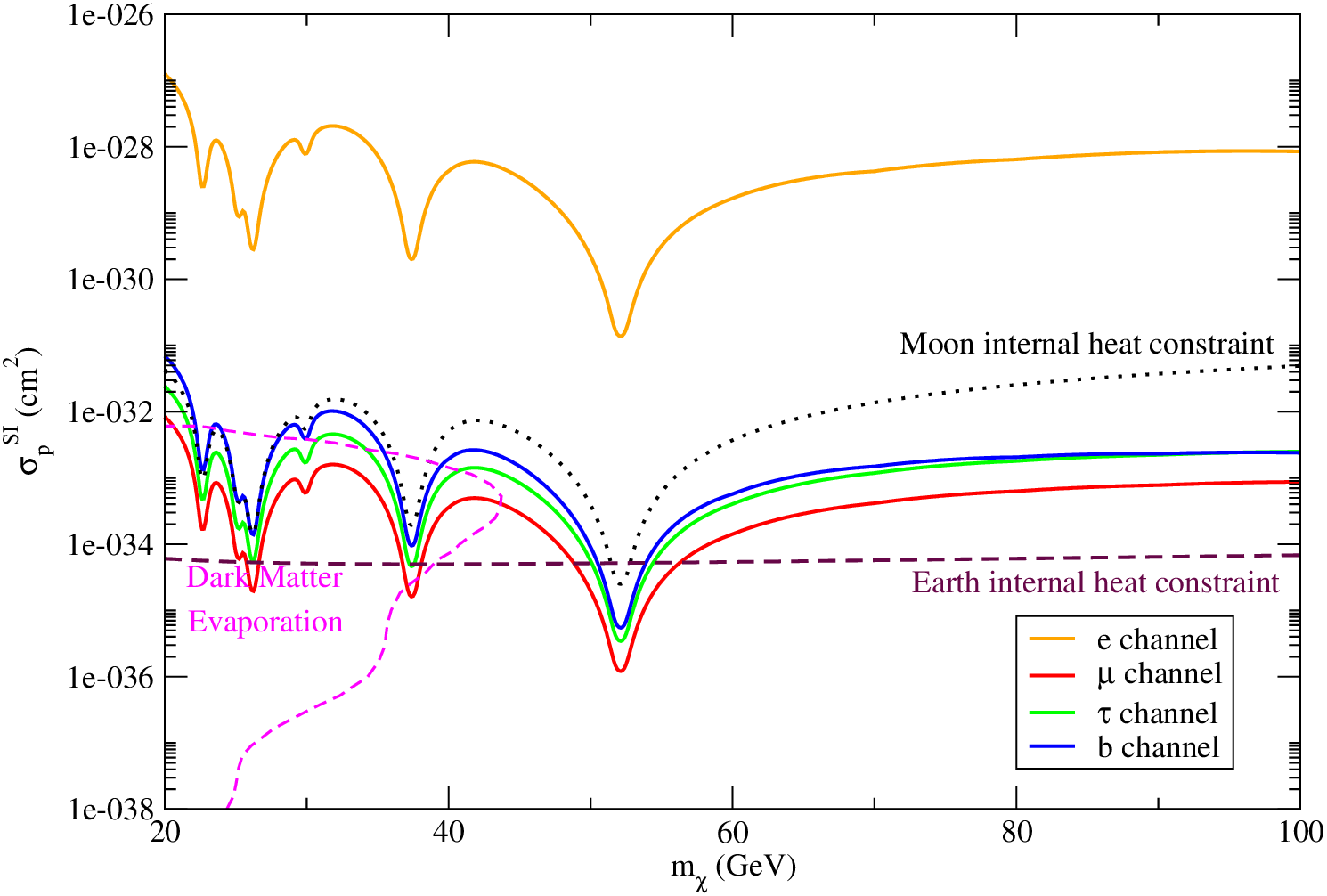} \newline
 \vskip 8mm
 \includegraphics[width=80mm]{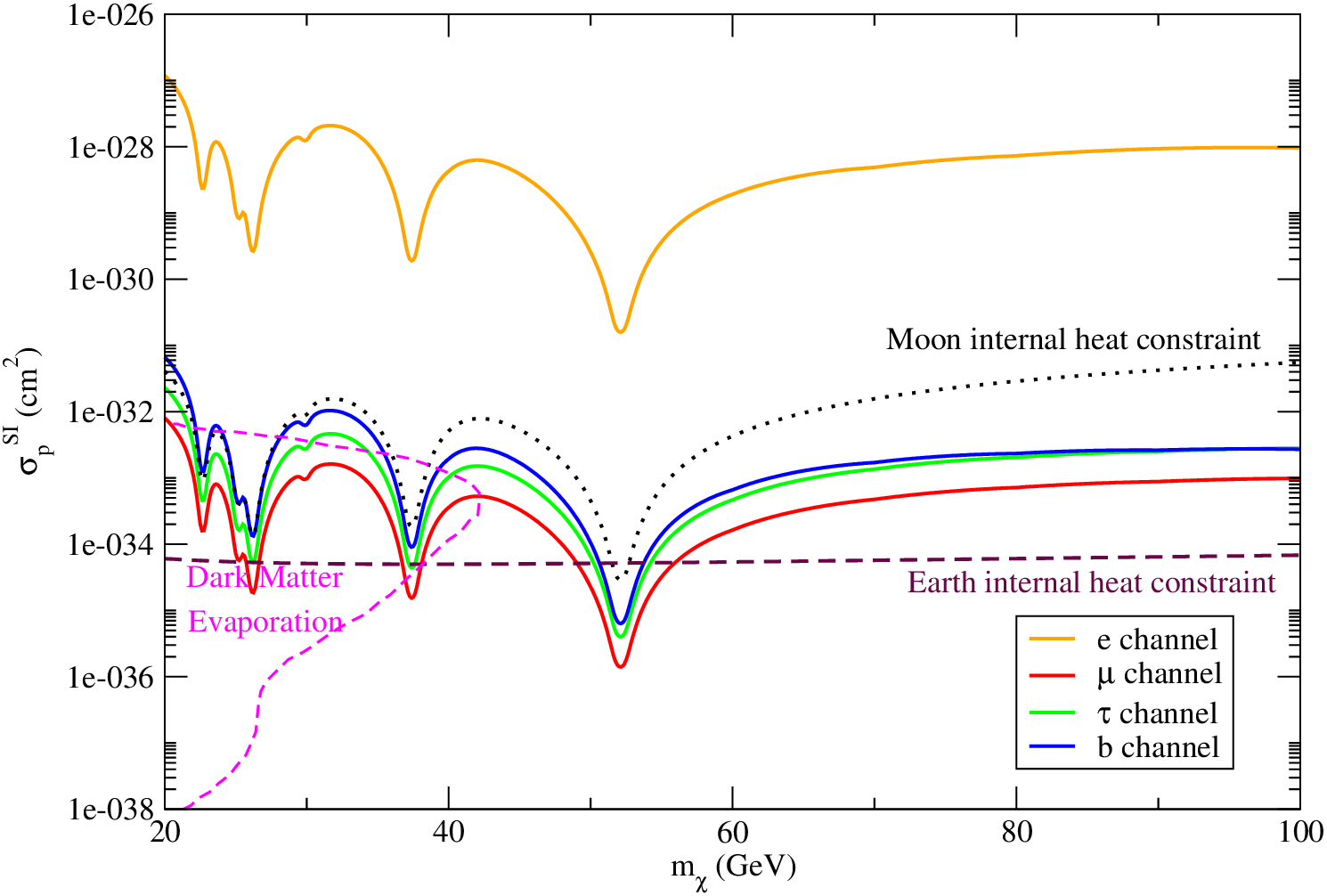}
\caption{The upper limits of $\sigma_p^{\rm SI}$ for MAX (upper figure) and MIN (lower figure) models, assuming dark matter annihilating via $e^+e^-$ (orange line), $\mu^+\mu^-$ (red line), $\tau^+\tau^-$ (green line) or $b\bar{b}$ (blue line) channels. The black dotted lines and the maroon dashed lines represent the upper limits obtained by the arguments of Moon internal heat \cite{Garani} and Earth internal heat \cite{Mack} respectively. The areas bounded by the pink dashed lines are the excluded regions due to dark matter evaporation \cite{Garani}. The neutrino data used are obtained from the Super Kamiokande detector and the calculation model in \cite{Honda}.}
\label{Fig2}
\end{figure}

\begin{figure}
 \includegraphics[width=80mm]{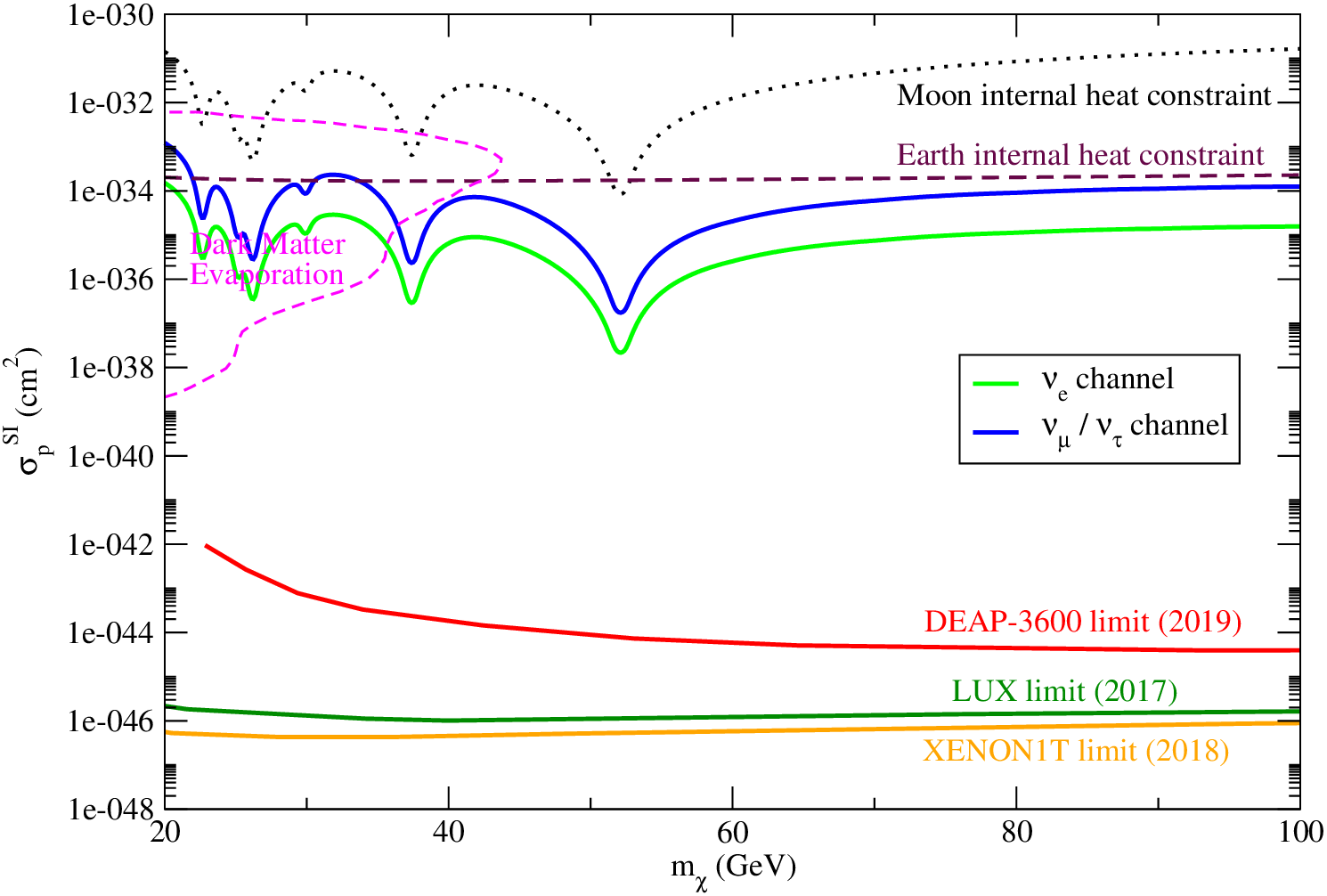} \newline
 \vskip 8mm
 \includegraphics[width=80mm]{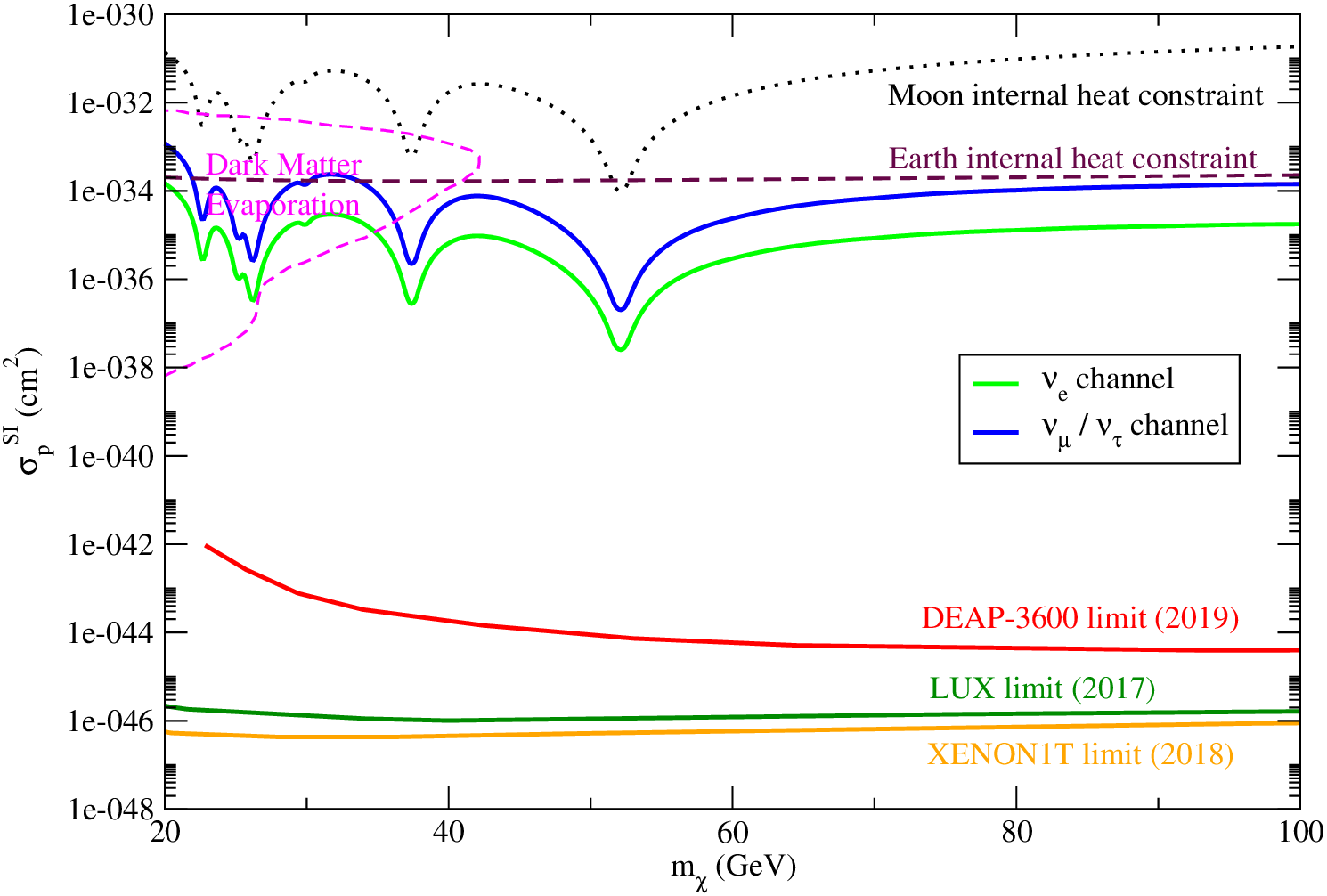}
\caption{The upper limits of $\sigma_p^{\rm SI}$ for MAX (upper figure) and MIN (lower figure) models, assuming dark matter annihilating via $\nu_e$ (green line), $\nu_{\mu}$ or $\nu_{\tau}$ (blue line) channels. The black dotted lines and the maroon dashed lines represent the upper limits obtained by the arguments of Moon internal heat \cite{Garani} and Earth internal heat \cite{Mack} respectively. The areas bounded by the pink dashed lines are the excluded regions due to dark matter evaporation \cite{Garani}. We also show the upper limits obtained by three direct-detection experiments (LUX \cite{Akerib}, XENON1T \cite{Aprile} and DEAP-3600 \cite{Yaguna}). The neutrino data used are obtained from the Super Kamiokande detector and the calculation model in \cite{Honda}.}
\label{Fig3}
\end{figure}

\begin{table}
\caption{Moon layers and elemental abundances for the MAX model \cite{Garani}.}
\begin{tabular}{ |c|c|c|c| }
 \hline\hline
 Radius & Density & Mass fraction & number of nuclei \\
 $[{\rm km}]$ & $[{\rm g/cm^3}]$ & $[\%]$ & $[\times 10^{46}]$ \\
 \hline\hline
    0-450    & 9   & Fe (95)               & Fe (3.52) \\
             &     & S (5)                 & S (0.32) \\
  \hline
    450-1680 & 3.2 & FeO (18) & O (88.0) \\
             &     & ${\rm Si_2O_3} (45)$ & Mg(8.3)\\
             &     &${\rm Al_2O_3} (14)$  & Al(10.3)\\
             &     & MgO (9)              & Si(32.5)\\
             &     & CaO (12)             & Ca (6.02)\\
             &     &                      & Fe (9.4)\\
  \hline
  1680-1737  & 2.9 & same                 & scaled by 0.097\\
  \hline \hline
\end{tabular}
\end{table}

\begin{table}
\caption{Moon layers and elemental abundances for the MIN model \cite{Garani}.}
\begin{tabular}{ |c|c|c|c| }
 \hline\hline
 Radius & Density & Mass fraction & number of nuclei \\
 $[{\rm km}]$ & $[{\rm g/cm^3}]$ & $[\%]$ & $[\times 10^{46}]$ \\
\hline\hline
    0-380    & 5   & Fe (95)               & Fe (1.1) \\
             &     & S (5)                 & S (0.10) \\
  \hline
    380-1697 & 3.4 & FeO (18) & O (97.1) \\
             &     & ${\rm Si_2O_3} (45)$ & Mg(9.2)\\
             &     &${\rm Al_2O_3} (14)$  & Al(11.3)\\
             &     & MgO (9)              & Si(35.8)\\
             &     & CaO (12)             & Ca(6.6)\\
             &     &                      & Fe (10.3)\\
  \hline
  1697-1737  & 2.4 & same                 & scaled by 0.051\\
  \hline \hline
\end{tabular}
\end{table}

\section{Discussion}
In this article, we present the calculation of the astrophysical upper limits of $\sigma_p^{\rm SI}$ using the neutrino data. We have got tighter constraints for $m_{\chi} \approx 48-56$ GeV for three popular annihilation channels ($\mu^+\mu^-$, $\tau^+\tau^-$ and $b\bar{b}$) and $m_{\chi} \ge 35$ GeV for the neutrino annihilation channels. The previous tightest constraints obtained are based on the arguments of the internal heat of the Moon and the Earth \cite{Garani,Mack,Kavanagh}. However, the maximum internal heat rates of the Moon and the Earth are highly model-dependent. For example, the heat flow models of the Earth's core give the residual heat flow output values from 2.3 TW to 21 TW \cite{Mack}. These constraints are obtained from our limited knowledge of heat flow inside the Earth's interior or Moon's interior. Nevertheless, our analysis applies the data of the GeV neutrino background flux (the atmospheric neutrino data), which are more reliable and were well-measured in the past two decades \cite{Honda,Gaisser}. Thus, our results can represent more reliable astrophysical constraints of $\sigma_p^{\rm SI}$.

Moreover, many recent studies are paying more attention to $m_{\chi} \approx 48-67$ GeV as this range of annihilating dark matter mass can simultaneously explain the gamma-ray excess and anti-proton excess in our Galaxy \cite{Daylan,Calore,Cholis}. This range can also explain some radio continuum data of galaxy clusters \cite{Chan3,Chan4}. Therefore, our tighter constraints on the range $m_{\chi} \approx 48-56$ GeV can provide some important complementary information of the particle dark matter. 

In fact, the observed atmospheric neutrino flux is larger than the real background neutrino flux \cite{Athar}. It is because some of the atmospheric neutrino flux is contributed by cosmic rays. Therefore, if we can obtain the real background neutrino flux in the future (e.g. detect from space), our constraints on $\sigma_p^{\rm SI}$ would be tighter. Furthermore, the neutrino flux emitted in dark matter annihilation inside the Moon has a specific direction (emitted from the Moon's solid angle). If we can observe the upper limit of the `lunar neutrino flux' in the future, our constraints on $\sigma_p^{\rm SI}$ would be much tighter because the `lunar neutrino flux' should be very small. However, this requires a very sensitive neutrino detector which has a very high resolution. 

\section{Acknowledgements}
This work was supported by a grant from the Research Grants Council of the Hong Kong Special Administrative Region, China (Project No. EdUHK 28300518).

\end{document}